\documentclass[fleqn,twoside]{article}
\usepackage{espcrc2}
\usepackage{graphicx}
\usepackage[figuresright]{rotating}
\usepackage{axodraw}

\newcommand{\AmS}{{\protect\the\textfont2
  A\kern-.1667em\lower.5ex\hbox{M}\kern-.125emS}}

\title{Charmonium Production in $e^+e^-$ Annihilation at $\sqrt{s}=10.6GeV$}

\author{Kuang-Ta Chao\address[add2]{Department of Physics, Peking University, Beijing 100871, People's Republic of China}
        and
        Li-Kun Hao\addressmark[add2]}

\begin{document}

\begin{abstract}

We review the charmonium production in $e^+e^-$ annihilation at
$\sqrt{s}=10.6GeV$. NRQCD predictions for the $J/\psi$ production
are compared with the recent measurements by BaBar and Belle. A
NRQCD calculation for the $D$-wave charmonium production is
reported to further test the color-octet mechanism and detect the
$J^{PC}=2^{--}$ state. The issue of double $c \bar c$ production
in $e^+e^-$ annihilation observed by Belle is also discussed.

\end{abstract}

\maketitle

\section{INTRODUCTION}

   The newly developed nonrelativistic QCD (NRQCD) factorization
formalism \cite{bbl} allows the infrared safe calculation of
inclusive heavy quarkonium production and decay rates. In the
NRQCD production mechanism, a heavy quark-antiquark pair can be
produced at short distances in a conventional color-singlet or a
color-octet state, and then evolves into an observed quarkonium
nonperturbatively. With this color-octet mechanism, one might
explain the Tevatron data on the surplus production of $J/\psi $
and $\psi ^{\prime }$ at large $p_T$, but puzzles about their
polarizatons still remain (for a review see \cite{kramer} and
references therein).

To further test the color octet mechanism, it may be interesting
to study the charmonium production in $e^+e^-$ annihilation. The
$J/\psi$ production in $e^+e^-$ annihilation has been investigated
within the color-singlet model \cite{cm1,cm2,cm3} and the
color-octet model \cite{om1,om2,ko}. The angular distribution and
energy distribution of color-singlet $J/\psi$ production at
$\sqrt{s}=10.6GeV$ have been discussed in \cite{cm3}. In
\cite{om1} it is noted that a clean signature of the color-octet
mechanism may be observed in the angular distribution of $J/\psi$
production near the end point region. In \cite{om2} contributions
of the color-octet as well as color-singlet to the $J/\psi$
production cross sections are calculated in a wide range of
$e^+e^-$ collider energies and it is found that with reasonable
choices of both color-singlet and octet matrix elements the color-
octet contribution will dominate. Moreover, the $J/\psi$
polarizations are predicted in \cite{ko}. Recently, BaBar
\cite{babar} and Belle \cite{belle} have measured the direct
$J/\psi$ production in continuum $e^+e^-$ annihilations at
$\sqrt{s}=10.6 GeV$. The total cross section and the angular
distribution seem to favor the NRQCD calculation over the
color-singlet model \cite{babar}, but some issues still remain.
The P-wave charmonium $\chi_{cJ}$ production in $e^+e^-$
annihilation has been discussed in \cite{xcprod}. The total
$\chi_{c1,2}$ cross sections are predicted to be dominated by the
color-octet process. In addition, as a further test of the
production mechanism, we will report a calculation for the
$D$-wave charmonium production in $e^+e^-$ annihilation, to which
the color-octet process will make substantial contributions
\cite{hao}, and these states may be detected in the future by
BaBar and Belle.

\section{$J/\psi$ PRODUCTION IN $e^+e^-$ ANNIHILATION}

The leading order color-singlet contributions to direct $J/\psi$
production include the following processes
\begin{eqnarray}
\label{eq1}
e^+e^-&\to&\gamma^*\to J/\psi gg,\\
\label{eq2}
e^+e^-&\to&\gamma^*\to J/\psi c\bar c,\\
\label{eq3} e^+e^-&\to&\gamma^*\to q\bar q g^*~~{\rm with}~~g^*\to
J/\psi gg.
\end{eqnarray}

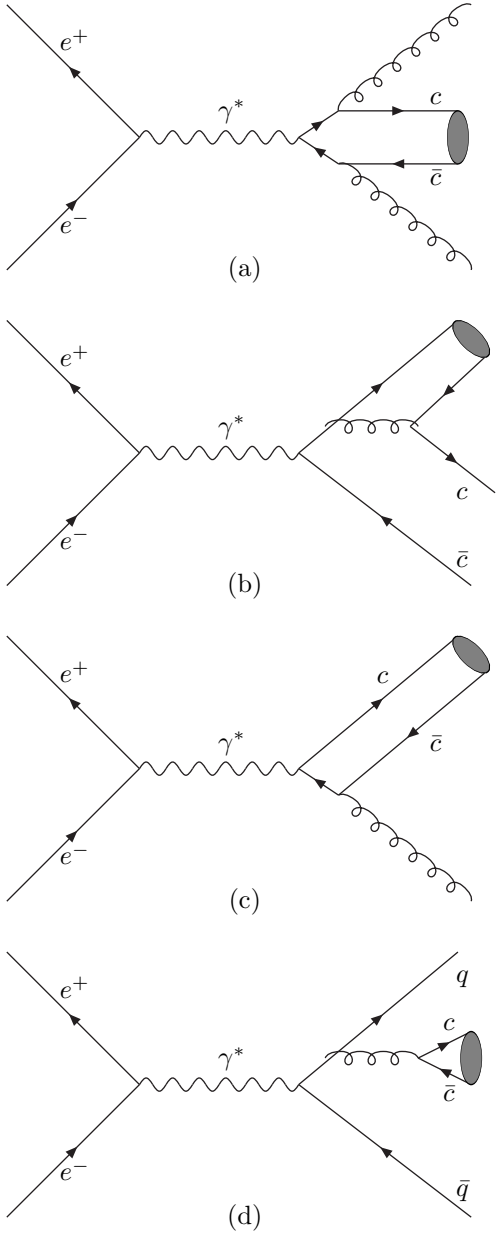
\begin{figure}
\centering
\begin{center}
\begin{picture}(200,110)(0,0)
\Text(90,10)[c]{(a)} \Text(80,70)[l]{$\gamma^*$}
\Text(20,28)[l]{$e^-$} \Text(20,97)[l]{$e^+$}
\Photon(50,60)(110,60){2.5}{6} \ArrowLine(0,10)(50,60)
\ArrowLine(50,60)(0,110) \ArrowLine(110,60)(125,70)
\ArrowLine(125,70)(170,70) \GOval(170,60)(10,4)(0){0.5}
\ArrowLine(170,50)(125,50) \ArrowLine(125,50)(110,60)
\Gluon(125,50)(175,10){2.5}{6} \Gluon(125,70)(175,110){2.5}{6}
\Text(160,75)[l]{$c$} \Text(160,45)[l]{$\bar {c}$}
\end{picture}\\
\vskip 3mm
\begin{picture}(200,110)(0,0)
\Text(90,10)[c]{(b)} \Text(80,70)[l]{$\gamma^*$}
\Text(20,28)[l]{$e^-$} \Text(20,97)[l]{$e^+$}
\Photon(50,60)(110,60){2.5}{6} \ArrowLine(0,10)(50,60)
\ArrowLine(50,60)(0,110) \ArrowLine(175,10)(110,60)
\Gluon(120,70)(155,70){2.5}{3} \ArrowLine(110,60)(170,110)
\ArrowLine(180,96)(152,70) \ArrowLine(152,70)(184,45)
\GOval(175,103)(9,4)(45){0.5} \Text(170,45)[l]{$c$}
\Text(170,20)[l]{$\bar {c}$}
\end{picture}\\
\vskip 3mm
\begin{picture}(200,110)(0,0)
\Text(90,10)[c]{(c)} \Text(80,70)[l]{$\gamma^*$}
\Text(20,28)[l]{$e^-$} \Text(20,97)[l]{$e^+$}
\Photon(50,60)(110,60){2.5}{6} \ArrowLine(0,10)(50,60)
\ArrowLine(50,60)(0,110) \ArrowLine(125,50)(110,60)
\ArrowLine(110,60)(170,110) \ArrowLine(180,96)(125,50)
\GOval(175,103)(9,4)(45){0.5} \Gluon(125,50)(175,10){2.5}{6}
\Text(140,95 )[l]{$c$} \Text(160,70)[l]{$\bar c$}
\end{picture}\\
\vskip 3mm
\begin{picture}(200,110)(0,0)
\Text(90,10)[c]{(d)} \Text(80,70)[l]{$\gamma^*$}
\Text(20,28)[l]{$e^-$} \Text(20,97)[l]{$e^+$}
\Photon(50,60)(110,60){2.5}{6} \ArrowLine(0,10)(50,60)
\ArrowLine(50,60)(0,110) \ArrowLine(175,10)(110,60)
\Gluon(120,70)(155,70){2.5}{3} \ArrowLine(110,60)(170,110)
\ArrowLine(155,70)(175,80) \ArrowLine(175,60)(155,70)
\GOval(175,70)(10,4)(0){0.5} \Text(170,100)[l]{$q$}
\Text(170,20)[l]{$\bar q$} \Text(165,83)[l]{$c$}
\Text(165,57)[l]{$\bar c$}
\end{picture}\\
\end{center}
\caption{The main Feynman diagrams for the charmonium production
in $e^+e^-$ annihilation. (a) and (b) are color-singlet processes;
while (c) and (d) are color-octet processes.}
\end{figure}

 Eq.(\ref{eq1}) is the gluon process, and Eq.(\ref{eq2}) is the
quark process (the charm quark fragmentation), and they are shown
respectively in Fig.1(a) and Fig.1(b). Eq.(\ref{eq3}) is the gluon
jet process with $q~=~u,~d,~s,~c$ quarks.

At $\sqrt{s}=10.6GeV$ the gluon process (1) dominates over the
other two color singlet processes. The process (3) is negligible
at low energies, but grows as the energy increases, and at high
enough energies ($\sqrt{s}\ge 250 GeV$) its contribution will
dominate over processes (1) and (2) \cite{om2}.

\begin{figure}
\centering
\begin{center}
\vspace{-3pc} \hspace{-4.5pc}
\includegraphics[width=22pc]{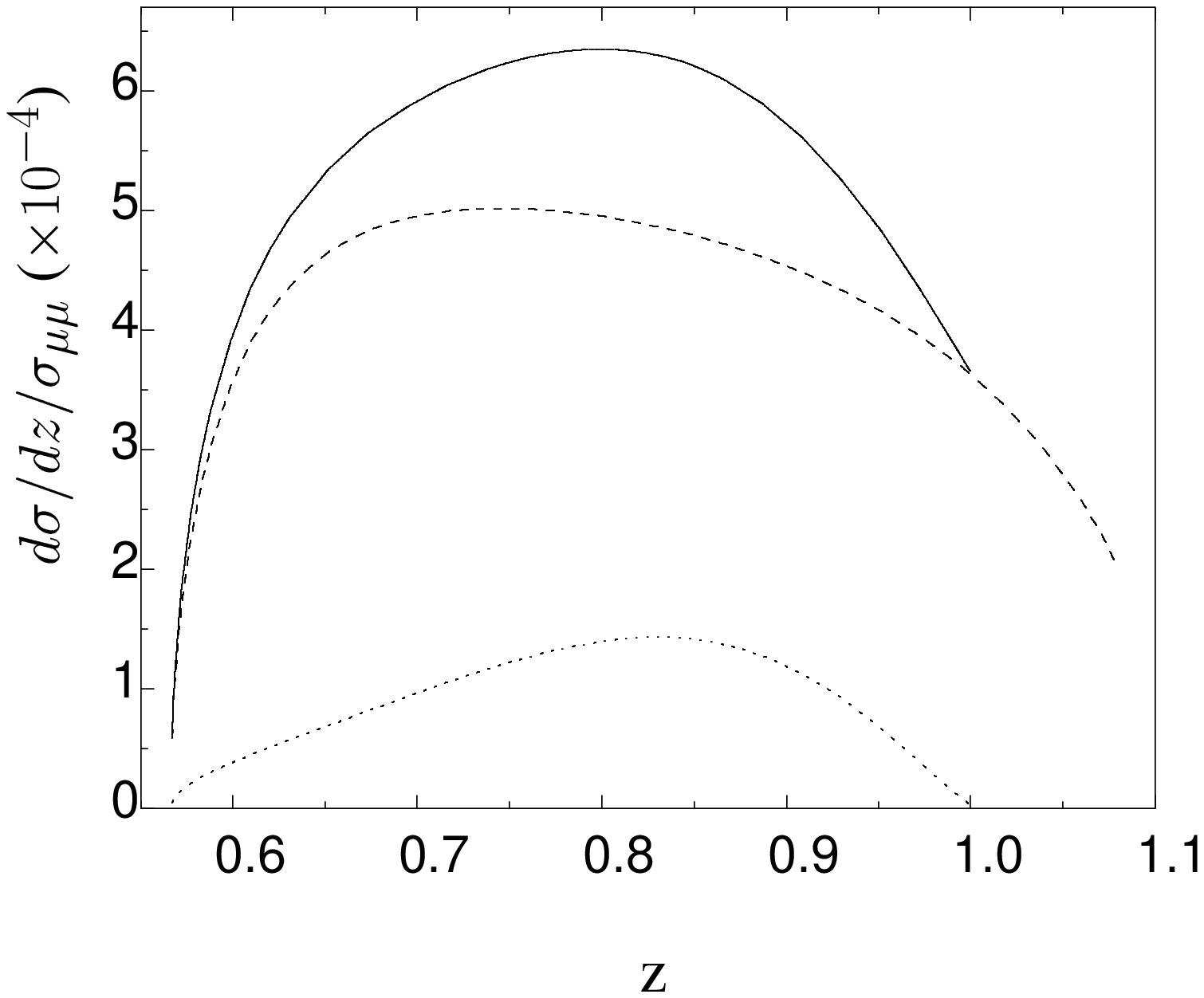}
\vspace{-15pc} \end{center} \caption{Color-singlet energy
distribution $d\sigma/dz$ as a function of
$z=\frac{2E_{\psi}}{\sqrt{s}}$ at $\sqrt{s}=10.6 GeV$. The
distributions are from the gluon process (1)~(dashed line), the
quark process (2)~(dotted line), along with the sum of them (solid
line). See Ref.\cite{om2}.}
\end{figure}

In Fig.2, We display the energy distributions of the gluon process
(1), the quark process (2), and their sum at $\sqrt{s}=10.6~GeV$.
The energy spectrum of these two processes are both flat and the
gluon process contribution is much larger than the quark process
(see \cite{om2}).

The leading order color-octet contributions to direct $J/\psi$
production in $e^+e^-$ annihilation include the following two
processes:
\begin{eqnarray}
e^+e^-&\to&\gamma^*\to g+c\bar c [\b{8},{^{2S+1}L_J}],  \\
e^+e^-&\to&\gamma^*\to q\bar q+c\bar c [\b{8},^3S_1],
\end{eqnarray}
as shown respectively in Fig.1(c) and Fig.1(d). Here ${
^{2S+1}L_J}$ denotes the states $^1S_0$ and $^3P_J$, $q$
represents the $u,~d,~s,~c$ quarks. At low energies the dominant
process is the process (4), whereas the process (5) will become
dominant at $\sqrt{s}>25GeV$.

As shown above, with a reasonable range of the color-singlet and
octet matrix elements, for the color-singlet cross section the
quark process (2) contributes 0.07-0.1 pb \cite{om2}, which is
consistent with \cite{cm3}, the gluon process (1) contributes
0.2-0.3 pb, and the total color-singlet cross section is 0.3-0.4
pb. The color-octet cross section contributes 0.6-0.7 pb.
Including both the color-singlet and octet contributions, we get a
total cross section of about 0.9-1.1 pb. BaBar gives the total
cross section $\sigma(e^+e^-\rightarrow J/\psi X)=2.52\pm 0.21\pm
0.21~pb$, and Belle gives $\sigma(e^+e^-\rightarrow J/\psi
X)=1.47\pm 0.10\pm 0.13~pb$. Their values are much larger than the
predicted color-singlet value of 0.3-0.4 pb. Moreover, for the
$J/\psi$ angular distribution parameter $A$, at high momentum
$p^*$, NRQCD predicts $0.6<A<1.0$ \cite{om1} while the
color-singlet model predicts $A\approx -0.8$, BaBar gives
$A=1.5\pm 0.6$ for $p^*>3.5GeV$, clearly favoring NRQCD.

However, there are some problems for the color-octet model. First,
the expected peak at the upper end point in the $J/\psi$ momentum
$p^*$ distribution in process (4) has not been observed. It is not
clear whether this could be due to the smearing effect of soft
gluon emission at the end point \cite{wolf}. Second, the $J/\psi$
polarization observed by BaBar and Belle is mainly longitudinal,
but the process (4) predicts the unpolarized $J/\psi$ \cite{ko}.
The color-singlet process (1) predicts longitudinally polarized
$J/\psi$ but it only gives a small portion of the cross section
($\sim$~0.3 pb). Again, the polarization problems remain as at the
Tevatron for the NRQCD color-octet mechanism \cite{cdf}. One of
the possible solutions for the $J/\psi$ polarization puzzle is
that the color $M1$ transitions, which will flip the spin of the
charm quark and give rise to longitudinal polarization, may not be
unimportant as compared with the $E1$ transitions in the evolution
of the charm quark pair.

\section{$\chi_{cJ}$ PRODUCTION IN $e^+e^-$ ANNIHILATION}

The P-wave charmonium $\chi_{cJ}$ production in $e^+e^-$
annihilation has been discussed in \cite{xcprod}. Because the
C-parity forbids the process $e^+e^-\rightarrow
c\bar{c}[^3P_J]gg$, the color-singlet processes are
$e^+e^-\rightarrow\chi_{cJ}+c\bar{c}$ and $e^+e^-\rightarrow
\chi_{cJ}+\gamma$, and the color-octet processes are
$e^+e^-\rightarrow (c\bar{c})_8[^1S_0,^3P_J]+g$. We get these
cross sections as $\sigma (e^+e^-\rightarrow
\chi_{cJ}\gamma)=0.001,0.012,0.005~pb$, $\sigma(e^+e^-\to
\chi_{cJ}+c\bar{c})=0.005,0.018,0.008~pb$, and
$\sigma_{octet}(\chi_{cJ})=0.022,0.067,0.112~pb$ for $J=0,1,2$
respectively. Then we get their total cross sections
\begin{equation}
\sigma (e^+e^-\rightarrow \chi_{cJ}X)=0.028,~0.097,~0.125~pb,
\end{equation}
which are consistent with \cite{xcprod}. The color-octet
contributions are larger than the color-singlet by a factor of
$2\sim 10$. These total cross sections are smaller than the upper
limit given by Belle.

\section{D-WAVE CHARMONIA $\delta_J$ PRODUCTION IN $e^+e^-$ ANNIHILATION}

In a recent study \cite{hao} we find that the spin triplet
$D$-wave charmonia $\delta_J (J=1,2,3)$ production in $e^+e^-$
annihilation at $\sqrt{s}=10.6$ GeV mainly comes from Fig.1(a) for
the color-singlet process, and Fig.1(c) for the color-octet
process. The contribution of color singlet process in Fig.1(b) to
the $\delta_1$ cross section is found to be smaller by a factor of
about 60 than that for the $J/\psi$ and therefore is negligible,
which is consistent with \cite{fragm}. Using the velocity scaling
rule and relating the $\delta_J$ color-octet matrix elements to
that of $\psi'$, we get the total cross sections
\begin{eqnarray}
\label{cs0} \nonumber
\sigma(\delta_1)=\sigma_1+\sigma_8\simeq0.027+0.016=0.043pb\\
\sigma(\delta_2)=\sigma_1+\sigma_8\simeq0.067+0.027=0.094pb
\end{eqnarray}

However, if we use a more radical choice of the $\delta_J$ color
octet matrix elements by relating them to that of $J/\psi$, we
would get much larger values for the $\delta_J$ production cross
sections, $\sigma(\delta_1)\simeq 0.16pb$ and
$\sigma(\delta_2)\simeq 0.29pb$. Therefore the measurement of
$\delta_J$ production in the future will be helpful to clarify the
color-octet mechanism and the $\delta_J$ color-octet matrix
elements.

In particular, for the $2^{--}$ D-wave charmonium $\delta_2$,
which is expected to be below the open charm $D \bar D^*$
threshold and forbidden to decay to $D \bar D$, its width is
estimated to be $300-400 KeV$ and its branching fraction of decay
mode $J/\psi\pi^+\pi^-$ is $B(\delta_2\rightarrow
J/\psi\pi^+\pi^-)\simeq 0.12$\cite{brr}, which is only smaller
than that of $B(\psi'\rightarrow J/\psi\pi^+\pi^-)=0.324\pm 0.026$
by a factor of 3. With more data available at the $B$ Factories in
the future, it will be possible to detect this $2^{--}$ D-wave
charmonium state. The $1^{--}$ D-wave state $\psi''(3770)$ could
also be detected via $\psi''\rightarrow D\bar{D}$ decay.

\section{PUZZLE OF DOUBLE $c \bar c$ PRODUCTION IN $e^+e^-$ ANNIHILATION}

Recently the Belle Collaboration has reported a measurement on the
double $c \bar c$ production in $e^+e^-$ annihilation at
$\sqrt{s}=10.6GeV$, and found that a very large fraction of the
produced $J/\psi$ is due to the double $c \bar c$ production in
$e^+e^-$ annihilation \cite{belle1}
\begin{eqnarray}
\nonumber \sigma(e^+e^-\rightarrow J/\psi c\bar
c)/\sigma(e^+e^-\rightarrow
J/\psi X)\\
= 0.59^{+0.15}_{-0.13}\pm 0.12,
\end{eqnarray}
which corresponds to $\sigma(e^+e^-\rightarrow J/\psi c\bar
c)\simeq 0.9 pb$.

This result is puzzling in terms of perturbative QCD calculations.
The color-singlet process (2) (the charm quark fragmentation, see
Fig.1(b)) will contribute only $\sim 0.1~pb$ to the cross section.
Moreover, the produced $J/\psi$ should be transversely polarized
in process (2)\cite{ko}, in contrast to the observed $J/\psi$
being mainly longitudinally polarized \cite{babar,belle}. The
color-octet process (5) may also contribute to the double $c \bar
c$ production with $q=c$, but again its cross section is too
small. Whether the QCD radiative correction to process (2) can
enhance the double $c \bar c$ production by a factor of $\sim 9$
remains interesting, but it may still be difficult to explain the
observed $J/\psi$ polarization. So, we intend to conclude that it
is very hard to explain the double $c \bar c$ production data
observed by Belle based on NRQCD or, more generally, on
perturbative QCD, and possible nonperturbative QCD effects have to
be considered.

In summary, we find that charmonium production in $e^+e^-$
annihilation at $\sqrt{s}=10.6GeV$ is interesting in testing the
heavy quarkonium production mechanism in NRQCD. We expect that
with more data collected at BaBar and Belle in the future, many
theoretical issues will be further clarified, and we will reach a
new step towards a better understanding for heavy quarkonium
physics in QCD.
\\

\noindent \textbf{Acknowledgments}

K.T.C. thanks the organizers of the BEACH 2002 Conference where
this work was presented. We also thank K.Y. Liu, C.F. Qiao, and F.
Yuan for their collaboration.


\begin{thebibliography}{99}

\bibitem{bbl}  G.T. Bodwin, L. Braaten, and G. P. Lepage, Phys. Rev. D51, 1125 (1995).

\bibitem{kramer} M. Kr\"amer, Nucl. Phys. {\bf B} (Proc. Suppl.) {\bf 93}, 176 (2001).

\bibitem{cm1} J.H. K\"uhn, J. Kaplan, and E.G.O. Sadiani, Nucl. Phys. {\bf B 157}, 125 (1979);
              C.-H. Chang, Nucl. Phys. {\bf B 172}, 425 (1980);
              B. Guberina, J.H. K\"uhn, R.D. Peccei and R. R\"uckl, {\it ibid}. {\bf B 174}, 317(1980);
              W.Y. Keung, Phys. Rev. {\bf D 23}, 2072 (1981);
              R. Baier and R. R\"uckl, Z. Phys. {\bf C 19}, 251 (1983);
              L. Clavelli, Phys. Rev. {\bf D26}, 1610 (1982);
              J.H. K\"uhn and H. Schneider, Z. Phys. {\bf C 11},253 (1981).

\bibitem{cm2} K. Hagiwara, A.D. Martin and W.J. Stirling, Phys. Lett. {\bf B267}, 527 (1991);
              V.M. Driesen, J.H. K\"uhn and E. Mirkes, Phys. Rev.{\bf D 49}, 3197 (1994).

\bibitem{cm3} P. Cho and K. Leibovich, Phys. Rev. {\bf D 54}, 6990 (1996).

\bibitem{om1} E. Braaten and Y.-Q. Chen, Phys. Rev. Lett. {\bf 76}, 730 (1996).

\bibitem{om2} F. Yuan, C.F. Qiao, and K.T. Chao, Phys. Rev. {\bf D 56}, 321 (1997); {\it ibid}, 1663 (1997).

\bibitem{ko} S. Baek, P. Ko, J. Lee, and H.S. Song, J. Korean Phys. Soc. {\bf 33}, 97 (1998).

\bibitem{babar} BaBar Collaboration, B. Aubert {\it et al}., Phys. Rev. Lett. {\bf 87},162002 (2001).

\bibitem{belle} Belle Collaboration, K. Abe {\it et al}., Phys. Rev. Lett. {\bf 88}, 052001(2002).

\bibitem{wolf} M. Beneke, G.A. Schuler, and S. Wolf, Phys. Rev. {\bf D 62}, 034004 (2000).


\bibitem{cdf} CDF Collaboration, T. Affolder {\it et al}., Phys. Rev. Lett. {\bf 85}, 2886 (2000).

\bibitem{xcprod} G.A. Schuler, Phys. Rev. {\bf D 58}, 017502 (1998).

\bibitem{hao} L.K. Hao, K.Y. Liu, and K.T. Chao, hep-ph/0206226.

\bibitem{fragm} K. Cheung, T.C. Yuan, Phys. Rev. {\bf D 53}, 3591 (1996).

\bibitem{brr} C. F. Qiao, F. Yuan, and K. T. Chao, Phys. Rev. {\bf D55}, 4001 (1997);
 {\bf D55}, 5437 (1997).

\bibitem{belle1} Belle Collaboration, K. Abe {\it et al}., hep-ex/0205104.


\end{thebibliography}
\end{document}